# A Bio-Inspired Whisker Sensor toward Underwater Flow Sensing in Darkness and Turbidity

1st Zheyi Hang
ZJUI Institute, International Campus
Zhejiang University
Haining, China
Zheyi.23@intl.zju.edu.cn

2nd Denghan Xiong
ZJUI Institute, International Campus
Zhejiang University
Haining, China
Denghan.22@intl.zju.edu.cn

3rd Pengbo Xie
ZJUI Institute, International Campus
Zhejiang University
Haining, China
Pengbo.24@intl.zju.edu.cn

4th Huan Hu*
ZJUI Institute, International Campus
Zhejiang University
Haining, China
huanhu@intl.zju.edu.cn

*Abstract*—Underwater flow sensing is critical for unmanned underwater vehicles (UUVs) and environmental monitoring, yet existing sensors often suffer from low responsiveness, high detection thresholds, and limited directional discrimination, along with complex packaging and poor long-term stability—especially for navigation and target perception in turbid and cluttered waters. Previous solutions based on traditional strain gauges with limited detection accuracy, or doped silicon sensors with limited detection height, have shown feasibility but still face challenges in scalability, robustness under harsh aquatic conditions, and calibration complexity. This work presents a bio-inspired whisker sensor that provides a balanced solution by embedding high-gauge-factor silicon strain gauges into a flexible PDMS base, mimicking seal whiskers to offer both high sensitivity and simplified packaging. The device exhibits a linear force–resistance response with a limit of detection of 0.27 mN, maintains stability after 10,000 loading cycles, and shows minimal offset drift of less than 2%, and shows frequency matching in underwater dipole tests with clear longitudinal and transverse spatial response patterns. These results indicate a robust and scalable route for underwater flow sensing on UUV platforms in practical deployments.

*Keywords-Bio-inspired sensor; Whisker sensor; Underwater flow sensing; Unmanned underwater vehicle; Silicon strain gauge*

## I. INTRODUCTION

Underwater flow sensing technology is crucial for the autonomous navigation, target recognition, and environmental monitoring of unmanned underwater vehicles (UUVs) [1-2]. In turbid or dark underwater environments, optical and acoustic sensing methods are often limited, whereas flow-based sensing approaches demonstrate unique advantages. In nature, marine mammals like seals can detect extremely weak water flow disturbances as low as 245 μm/s using their specialized whisker structures. They can precisely track vortex wakes generated by fish swimming, even when these wakes have persisted for several minutes [3-4]. This exceptional sensing capability arises from the combination of highly sensitive neural receptors at the whisker base together with the unique asymmetric wavy structure of the whiskers, which effectively suppresses vortex-induced vibrations (VIV), enabling the extraction of target signals in complex flow fields [5-6].

Inspired by such biological mechanisms, biomimetic whisker flow sensors have become a research hotspot in underwater sensing [7-8]. Currently, researchers worldwide have developed biomimetic whisker sensors based on various principles, including piezoresistive [9], piezoelectric [10], capacitive [11], and other approaches. However, existing approaches still face common challenges [12-13]: sensors based on traditional strain gauges offer low cost but limited detection accuracy, while doped silicon sensors provide higher sensitivity but suffer from limited detection height for hair sensing. Their manufacturing processes are complex, costly, and difficult to package, with insufficient long-term reliability and robustness in harsh underwater environments [2]. Moreover, most studies remain confined to single-point sensing, struggling to replicate the spatial flow field perception capabilities of biological whisker arrays, and their calibration processes are often complex [14-15].

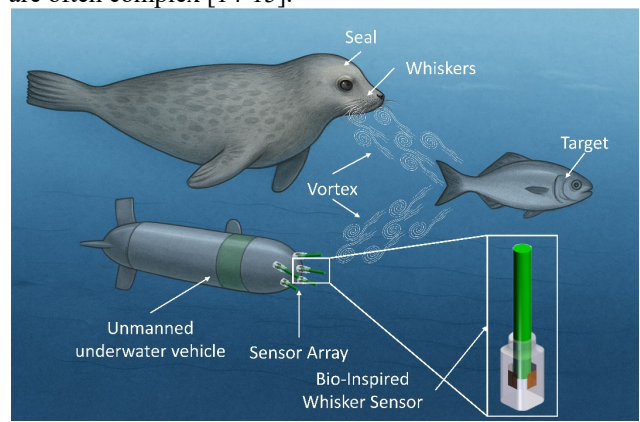

Figure 1. Cartoon Illustration of the Bio-Inspired Whisker Sensor Array Principle.

To address these issues, this paper presents a balanced solution that integrates the high sensitivity of silicon strain gauges with the protective flexibility of a PDMS base. We designed and fabricated a biomimetic whisker sensor whose core innovation lies in embedding highly sensitive silicon strain gauges into a flexible PDMS base, integrated with a 3D-printed whisker. Unlike active acoustic systems, this sensor relies on passive flow sensing, avoiding additional signal emission while still achieving both high sensitivity and robust environmental adaptability, with manageable manufacturing cost and scalability.

To validate the sensor's performance, we conducted a series of experimental studies. First, mechanical calibration tests confirmed excellent linearity, repeatability, and stability in its force-resistance response. Fatigue cycling tests (exceeding 10,000 cycles) demonstrated outstanding long-term reliability. Subsequently, underwater dipole experiments successfully reproduced the sensor's response to specific frequency flow field excitation, revealing distinct longitudinal and transverse spatial response patterns. These results demonstrate not only the sensor's superior performance but also its scalability for large-scale array deployment, offering a viable technical pathway for robust, low-cost underwater flow sensing and target tracking on UUV platforms.

## II. Sensor Design and Fabrication

### A. Sensor Design

The schematic diagram of the bio-inspired whisker sensor is presented in Figure 2(a). Its core consists of three components: PDMS base, silicon strain gauge, and Artificial Whisker. The PDMS base is designed as a square prism with a base area of 10 mm × 10 mm and a height of 25 mm. At its top center, a circular hole with a diameter of 5 mm and a depth of 15 mm is designed to secure the Artificial Whisker. A key design feature is that the bottom depth of this hole is engineered to align precisely with the centerline of the silicon strain gauge. This layout ensures that when the Artificial Whisker bends under force, the maximum strain generated at its roots is effectively transmitted to the sensitive region of the silicon strain gauge, enhancing the piezoresistive effect and thereby achieving optimal sensor sensitivity.

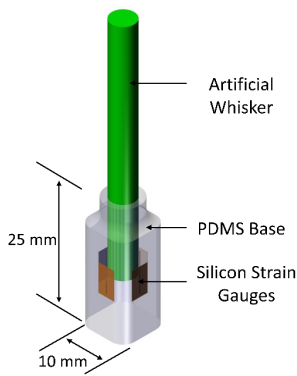

Figure 2. Schematic illustration and prototype of the bio-inspired whisker sensor. (a) 3D schematic model of the sensor, (b) Photograph of the sensor prototype.

Specialized grooves for embedding the silicon strain gauge are designed on the side walls of the PDMS base. These grooves have a depth of 1.25 mm and dimensions of 13 mm × 4 mm. The selected silicon strain gauge measures 6.0 mm × 3.5 mm × 0.22 mm (length × width × thickness). After two PDMS curing cycles, the silicon strain gauges are firmly embedded within the PDMS base. Their surfaces remain approximately 1 mm from the device's outer surface, ensuring both insulation protection and effective strain transfer. Both leads of the silicon strain gauge extend downward from the column base and are waterproofed. The Artificial Whiskers are straight cylinders with a diameter of 5 mm and a length of 100 mm, fabricated via 3D printing using PLA material, which exhibits significantly greater material stiffness than the PDMS base. The final prototype is shown in Figure 2(b).

### B. Sensor Fabrication

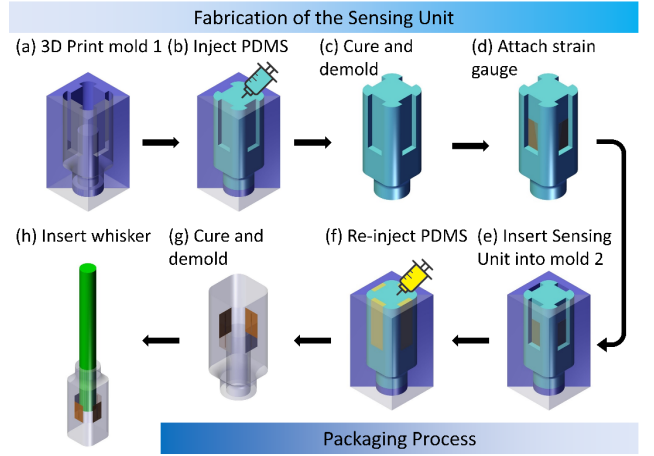

Figure 3. Fabrication process of the sensor. (a) 3D Print mold 1, (b) Inject PDMS, (c) Cure and demold, (d) Attach strain gauge, (e) Insert Sensing Unit into mold 2, (f) Re-inject PDMS, (g) Cure and demold, (h) Insert whisker.

The sensor fabrication process is illustrated in Figure 3. The first step involves mold design and printing (Figure 3(a)). Two 3D-printed molds, Mold 1 and Mold 2, each 14 mm × 14 mm × 27 mm, were fabricated with PLA using a Bambu Lab X1D FDM printer. Mold 1 had protrusions on its side walls to create recesses for silicon strain gauges, while Mold 2 had smooth walls; both were diagonally split for easy demolding after PDMS curing. As shown in Figure 3(b), Sylgard 184 prepolymer and curing agent were mixed at 10:1, vacuum-degassed, poured into Mold 1, and cured naturally at room temperature for 24 h to obtain a PDMS base blank with sidewall grooves. Silicon strain gauges were then mounted (Figure 3(d)) by applying silicone adhesive into the grooves, positioning each gauge with tweezers, and pressing gently to ensure contact, so that all four side walls were embedded. This PDMS base with mounted gauges was used as the sensing unit. For secondary encapsulation, the sensing unit was placed into Mold 2 (Figure 3(e)), leaving voids along the grooves. Degassed PDMS was injected (Figure 3(f)), cured again for 24 h, and then demolded. The sample was subsequently heated at 70 °C for 2 h to complete PDMS curing, strengthen the structure, and ensure sealing performance, fully encapsulating the strain gauges within the PDMS base (Figure 3(g)). Finally, a 3D-printed whisker, 5 mm in diameter and 100 mm in length, was inserted into a cylindrical hole (5 mm diameter, 15 mm depth) at the center of the PDMS base, precisely aligned with the strain gauge center (Figure 3(h)), completing fabrication of the bio-inspired whisker sensor.

## III. STATIC AND REPETITIVE BENDING TESTS

### A. Static Bending Characterization and Calibration

Each silicon strain gauge is connected in a ¼-bridge Wheatstone configuration. The bridge output is amplified by an AD620 amplifier (gain = 23.5) and digitized by an NI data-acquisition device at 6.25 kHz. To quantify the static bending response in air, we perform quasi-static lateral loading at the Artificial Whisker tip using a flat-headed push rod oriented perpendicular to the whisker axis. The applied force was stepped from 0 to 0.18 N in 0.02 N increments, and data at each force level were recorded after signal stabilization. Ten complete loading–unloading cycles were conducted to assess repeatability.

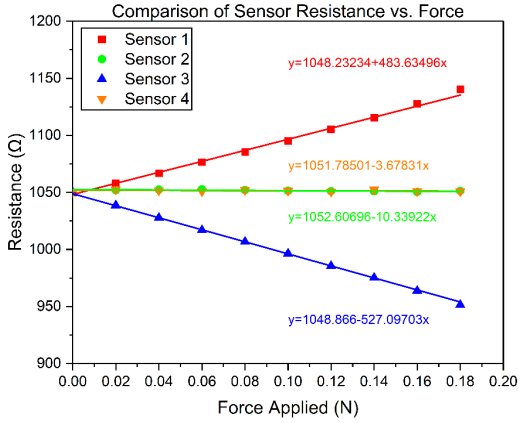

Figure 4. Comparison of Sensor Resistance vs. Force.

The resistance change of each embedded silicon strain gauge was well described by a linear model,

$$R(F) = R_0 + kF \quad (1)$$

With representative results shown in Figure 2(a), at the maximum load of 0.18 N the tip displacement was 1.5 mm, corresponding to an estimated deflection angle of about 10°. This upper limit ensured adequate signal magnitude while keeping the device within the elastic regime. The pair of gauges aligned with the principal bending axis, Sensor 1 and Sensor 3, showed opposite-sign but closely matched sensitivities, with +483.63 Ω/N for Sensor 1 and −527.10 Ω/N for Sensor 3. These opposite signs reflect tensile versus compressive strain, while the comparable magnitudes confirm packaging symmetry and precise load alignment. In contrast, the orthogonal gauges, Sensor 2 and Sensor 4, produced near-zero responses with sensitivities of −10.34 Ω/N and −3.68 Ω/N, demonstrating effective suppression of off-axis coupling and enabling differential bridge readout with temperature-drift compensation. At the maximum load, resistance changes reached +87.1 Ω for Sensor 1 and −94.9 Ω for Sensor 3, corresponding to relative variations of 8.3% and 9.0%.

The single-channel force limit of detection (LOD) was estimated using

$$LOD = \frac{3 \times \sigma_y}{S} \quad (2)$$

where $\sigma_y$ is the standard deviation of the resistance noise under zero load and S is the fitted sensitivity (slope) for the channel. With $\sigma_y = 0.044\,\Omega$, the LOD is $2.69 \times 10^{-4}$ N (0.27 mN).

### B. Cyclic Repeatability and Stability

A reciprocating machine drove the Artificial Whisker with a 1.5 cm stroke. We focused on the 1.5 Hz, 10,000-cycle condition, with electrical sampling at 100 Hz. Figure 5 shows a representative time window in which the two gauges aligned with the principal bending axis (Sensors 1 and 3) exhibit phase-opposed tensile/compressive responses with closely matched amplitudes in each cycle, consistent with the static-bending results.

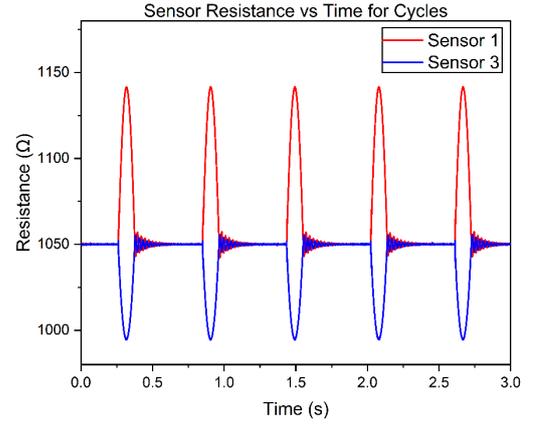

Figure 5. Sensor Resistance vs. Time for Cycles.

The 10,000-cycle record was segmented by cycle, and two metrics were tracked: the per-cycle maximum of Sensor 1 and the per-cycle minimum of Sensor 3 as shown in Figure 6. Both trajectories showed only slow absolute drift relative to the nominal 1050 Ω baseline. The cumulative offset of the Sensor 1 maximum was 1.9% and that of the Sensor 3 minimum was 1.0%, corresponding to drift rates of 2.0 ppm per cycle and 1.1 ppm per cycle, respectively. No distortion, abrupt change, or phase reversal was observed over 10,000 cycles, indicating no fatigue-related degradation in the silicon strain gauges, lead interconnects, or PDMS base encapsulation under this loading condition.

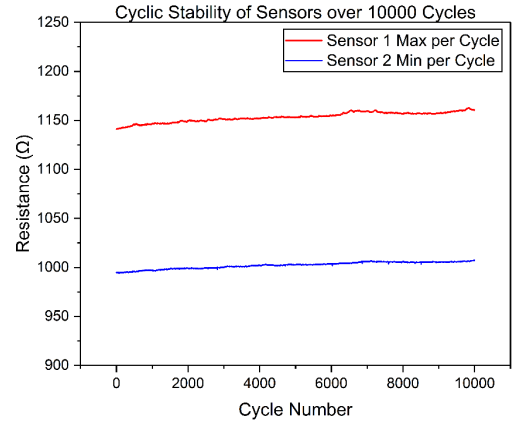

Figure 6. Cyclic Stability of Sensors over 10000 Cycles.

Through the static bending test and long-term cyclic fatigue experiments, it is evident that the sensor can accurately measure both the amplitude and direction of whisker bending. Moreover, under different frequencies and cycle numbers, the sensor exhibits stable output characteristics, demonstrating good linearity, high repeatability, and reliable fatigue durability.

## IV. UNDERWATER FLOW SENSING EXPERIMENTS

### A. Experimental Setup and Data Acquisition

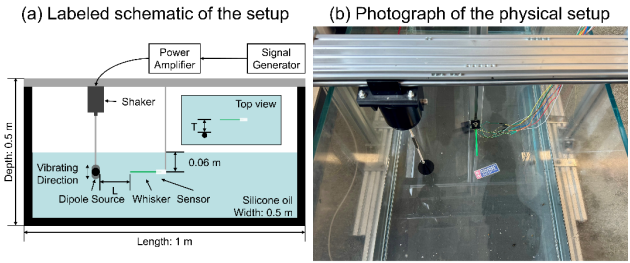

Figure 7. Underwater Flow Experiment Setup. (a) Labeled schematic of the setup, (b) Photograph of the physical setup.

As shown in Figure 7, we built a benchtop platform to evaluate the sensor in a viscous fluid. A transparent tank (100 cm × 50 cm × 50 cm) was filled with silicone oil. A signal generator and power amplifier drove an electrodynamic shaker mounted on the tank lid; a stainless-steel rod attached to the shaker moved a rigid ball in periodic reciprocation, producing a canonical dipole flow field. The sensor was fixed near the tank center. The source–sensor geometry was defined by the longitudinal distance L along the drive axis and the transverse offset T orthogonal to it. To ensure clear orientation, the excitation axis was kept perpendicular to the whisker axis.

Each silicon strain gauge was wired in a ¼-bridge Wheatstone configuration. Outputs were conditioned by an AD620 amplifier (gain = 166) and digitized across four channels by an NI data-acquisition device at 6.25 kHz.

Signal processing for each (L, T, f) condition included anti-alias filtering, decimation to 100 Hz, segmentation into fixed windows, removal of mean drift, and FFT to obtain amplitude spectra. The response metric was the spectral magnitude at the commanded frequency; harmonics were monitored but excluded from analysis. Each data point reports the mean of 10 repeated trials, with error bars showing standard deviation.

### B. Frequency Response Characterization

At fixed geometry (L = 20 mm, T = 0 mm), we swept the commanded frequency from 1 to 50 Hz and extracted the detected frequency from the FFT peak. Figure 8 shows the detected versus commanded frequency for all channels; the traces overlap, confirming excellent synchronization.

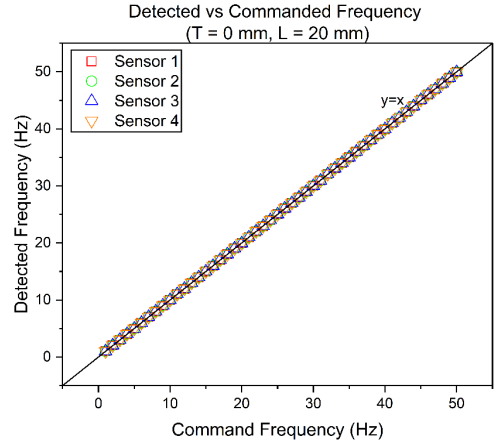

Figure 8. Detected vs Commanded Frequency.

A linear fit yielded a coefficient of determination $R^2$ of 0.999992, indicating a slope deviating only 0.063% from unity and a near-zero intercept. Across the band, absolute frequency error never exceeded 0.09 Hz, with median and mean absolute errors of 0.03 Hz and 0.033 Hz, and a root mean square error of 0.040 Hz. These results demonstrate highly accurate frequency tracking across 1–50 Hz with negligible discrepancies between channels.

### C. Response to Longitudinal Distance Variation

Figure 9 characterizes the sensor's response to a dipole source moving away along the longitudinal axis (T = 0 mm) at 10 Hz. The primary observation is an apparent decrease in signal amplitude across all four channels as the distance (L) increases, which aligns with the expected attenuation of a flow field.

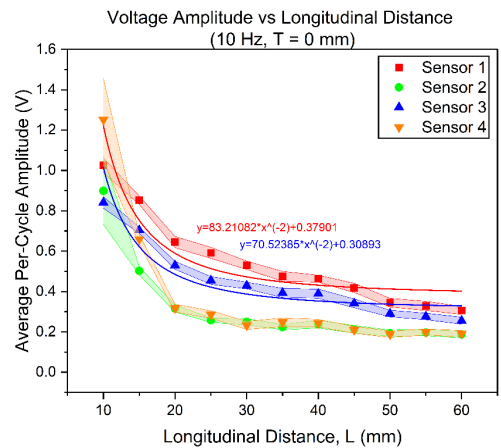

Figure 9. Voltage Amplitude vs Longitudinal Distance.

A more insightful finding lies in the relative amplitudes between the channels. In this specific configuration, where the source moves directly towards and away from the sensor, the signals from Sensor 1 and Sensor 3 are consistently stronger than those from Sensor 2 and Sensor 4. This indicates that the sensor's housing deforms more easily in the direction aligned with Sensor 1 and 3 when the flow stimulus comes from the front. In other words, the sensor is not equally sensitive in all directions; it possesses a directional preference.

The amplitude decays most rapidly within the first 20 mm (the near-field), after which the decrease slows, approaching a noise floor beyond 40 mm. This transition defines the effective operational range for distance estimation under these conditions to be approximately 40-50 mm. The key conclusion is that the sensor can not only detect the presence of a source through its amplitude response but also infer the source's movement direction relative to its own inherent sensitivity axes. The specific axis of highest sensitivity depends on the direction of the incoming flow, a property that is further explored in the transverse offset analysis.

*D. Response to Transverse Offset Variation*

The sensor's spatial selectivity was further tested by varying transverse offset T at fixed L = 20 mm and f = 10 Hz (Figure 10). The amplitude peaked when the source was aligned with the sensor center and decayed symmetrically as |T| increased, reaching the noise floor at about 15–20 cm. This profile matched the expected dipole distribution.

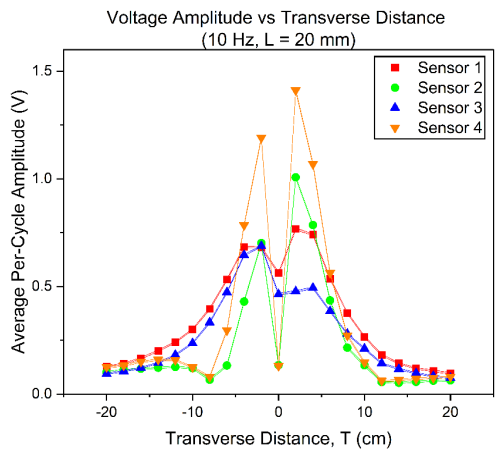

Figure 10. Voltage Amplitude vs Transverse Distance.

Distinct response magnitudes were observed between axes. Channels aligned with the transverse direction showed higher amplitudes than those along the longitudinal direction. At T = 0 cm, Sensor 4 reached about 1.41 V compared to 0.56 V for Sensor 1, indicating highest sensitivity to flow gradients when the source is displaced transversely. Consequently, Sensor 2 and Sensor 4 act as the primary elements for detecting transverse motion.

The profiles were symmetrical about T = 0 cm, confirming fabrication consistency and flow field stability. The narrow peaks, with significant amplitude decrease within about 2 cm, indicate sharp spatial selectivity, essential for high-resolution localization and direction discrimination.

Experimental results demonstrate that the sensor can detect both the distance and direction of an underwater target. The longitudinal test allows for distance estimation, while the transverse test provides directional information. This orthogonal sensing capability confirms the sensor's utility for practical applications such as underwater trajectory tracking.

## V. CONCLUSION

This work presents a bio-inspired whisker sensor as a balanced solution for underwater flow sensing. Experimental results confirm its excellent linearity, durability over 10,000 cycles, and accurate frequency tracking capability. The sensor successfully decodes both distance and direction of hydrodynamic sources through distinct longitudinal and transverse response patterns. These results highlight the sensor's promise for wake tracking and positioning tasks in underwater environments. Future work will reduce noise and optimize array topology toward robust trail detection and underwater localization.

ACKNOWLEDGMENT

This work was supported by ZJUI and led by Principal Supervisor Prof. Huan Hu.